\RequirePackage{fix-cm}
\documentclass[fleqn,usenatbib]{mnras}

\usepackage{newtxtext,newtxmath}
\usepackage{url}

\usepackage[T1]{fontenc}

\DeclareRobustCommand{\VAN}[3]{#2}
\let\VANthebibliography\thebibliography
\def\thebibliography{\DeclareRobustCommand{\VAN}[3]{##3}\VANthebibliography}

\usepackage{graphicx}	
\usepackage{amsmath}	
\usepackage[dvipsnames]{xcolor}

\newcommand{\mx}[0]{\mathrm{mx}}
\newcommand{\mxo}[0]{\mathrm{mx0}}

\defcitealias{Errani2021}{EN21}
\defcitealias{Errani2022}{E+22}

\newcommand{\msun}{$\rm M_\odot$}

\title[Ultra-faint dwarfs to giant ellipticals in galaxy clusters]  {The abundance and radial distribution of faint and ultra-faint dwarfs  in galaxy clusters}

 \author[Sadhu et al.]{
 Pradyumna Sadhu,$^{1}$\thanks{E-mail: psadh003@ucr.edu}
 Laura V. Sales,$^{1}$
 Julio F. Navarro,$^{4}$
 Rapha\"{e}l Errani,$^{2, 3}$
 Jose Benavides$^{1}$,
 \newauthor
 and Eric W. Peng$^{5}$
 \\
 $^{1}$ Department of Physics and Astronomy, University of California, Riverside, CA, 92521, USA \\
 $^{2}$ McWilliams Center for Cosmology, Department of Physics, Carnegie Mellon University, Pittsburgh, PA 15213, USA\\
 $^{3}$ Universit\'{e} de Strasbourg, CNRS, Observatoire Astronomique de Strasbourg, UMR 7550, F-67000 Strasbourg, France\\ 
 $^{4}$ Department of Physics and Astronomy, University of Victoria, Victoria, BC V8P 5C2, Canada \\
 $^{5}$ NSF NOIRLab, 950 N. Cherry Ave., Tucson, AZ 85719, USA 
 }

\date{Accepted XXX. Received YYY; in original form ZZZ}

\pubyear{2026}

\begin{document}
\label{firstpage}
\pagerange{\pageref{firstpage}--\pageref{lastpage}}
\maketitle

\begin{abstract}
Cosmological simulations of galaxy clusters are unable to resolve dwarf galaxies due to limited numerical resolution which drives the artificial disruption of dark matter substructures.  We address these limitations by combining the results of the cosmological hydrodynamical simulation TNG50 in $\Lambda$CDM with an empirical model of tidal evolution of cluster galaxies calibrated using high-resolution idealized N-body simulations. Applied to the three most massive clusters in TNG50, our model allows us to study the stellar mass and radial distribution of dwarfs well below the formal resolution limit of the parent simulation. We find that, at $z=0$, clusters with virial mass $M_{200} \sim 10^{14}$~\msun\; host a vast population of dwarf galaxies within the virial radius, amounting to $2000$-$7000$ systems with $M_* > 100$~\msun. Taken together, these satellites follow a radial distribution that matches the underlying dark matter profile of the host. However, applying a minimum mass or luminosity threshold for detection, as  expected in observational studies, tends to exclude the most heavily-stripped objects, which tend to populate the inner regions. Future surveys targetting ultra-faint galaxies in group and cluster environments, such as those made possible by the Euclid, Rubin, or Roman telescopes, will be fundamental to refute or confirm this prediction.
\end{abstract}




\section{Introduction}
\label{sec:intro}

Satellite galaxies are expected to inhabit the subhalos that have survived the hierarchical assembly of their host galaxy halo. In $\Lambda$CDM, the subhalo mass function follows a steep function that increases monotonically with decreasing subhalo mass ($m_{\rm sub}$): $dN/dm_{\rm sub} \propto m_{\rm sub}^{-1.9}$ \citep{Gao2004,Springel2008}. This implies that dwarf systems should dominate by number the satellite population of luminous galaxies or of galaxy clusters, making them an attractive target for observational and theoretical studies.

The galaxy population of galaxy clusters, in particular, provides a useful testbed of these theoretical expectations. One advantage is simple statistics. For example, while the Milky Way hosts about a dozen "classical" dwarf satellites ($M_* > 10^5\,$\msun), more massive environments should harbor a much larger population of such dwarfs; perhaps $500+$ dwarfs in the case of the Fornax cluster \citep{Venhola2019} or $\sim 300+$ just in the core (i.e., inside $\sim 300$ kpc) of the Virgo cluster \citep{Ferrarese2016}. Pushing observations into the ultra-faint (i.e., $M_*<10^5\, M_\odot$) regime should presumably lead to the discovery of thousands of faint satellites in such systems.   

However, a number of factors hinder definitive theoretical predictions of the  faint galaxy population. First, the dynamic range necessary to simulate dwarf galaxies in  clusters is much larger than in the field or in lower mass systems such as Milky Way (MW) mass galaxies. 

A second problem is that the number of surviving satellites in cosmological N-body simulations is affected by the artificial  disruption of subhalos due to limited force and spatial resolution. This effect is expected to bias the predicted  satellite population to lower numbers and less concentrated radial distributions \citep[e.g., ][]{vandenBosch2018b, Green2021, Santos-Santos2024}.
The artificial satellite disruption is expected to be secondary to limitations from numerical resolution itself \citep{Green2021}. However, inaccuracies due to both factors increase systematically for low mass satellites (poorer particle sampling) that orbit in the deeper regions of the gravitational potential \citep{vandenBosch2018b}. These limitations conspire to make the predictions for number, mass and radial distribution of ultra-faint-bearing subhalos particularly impacted. 

In addition, taking into account the baryonic physics relevant to star formation, feedback and gas transport -- necessary to simulate galaxy formation in cosmological N-body simulations-- incurs a high computational cost. As a result, state-of-the-art cosmological hydrodynamical simulations of representative volumes of the universe reach mass-per-baryonic particle of the order $10^5 \rm{-} 10^7$\msun\; \citep[e.g., ][]{Pillepich2018,Zhou2025,Schaye2025}, limiting the analysis of low mass galaxies to objects more massive than  $M_* \geq 10^7-10^9$\msun. As a result, even our most sophisticated simulations of group and cluster environments partially  miss the faint population and fully miss the ultra-faint regime. 

These limitations, however, may be partially alleviated by using idealized experiments of tidal evolution of single satellites to correct for artificial disruption and to account more faithfully for tidal effects on both the dark matter and the stellar components of a galaxy \citep{Hayashi2003, Kazantzidis2004}. 

For dark matter-dominated galaxies embedded in a cuspy dark matter subhalo that follows a Navarro, Frenk and White profile \citep[][NFW hereafter]{Navarro1996}, the tidal evolution of a subhalo can be parametrized in terms of the remaining bound mass after infall, enabling a description of the tidal remnant at several stages during their evolution  \citep{Penarrubia2008, Errani2015, Green2019, Errani2021}. 
More recently, \cite{Errani2022} has extended this formalism, allowing for a more detailed treatment that considers the tidal evolution not only of the dark matter but also of the stellar component. The latter, in particular, depends mainly on the initial energy distribution of the stars, as well as on their radial segregation relative to the dark matter.

The primary goal of this paper is to present a model which grafts the tidal evolution model of \cite{Errani2021, Errani2022} into the TNG50 cosmological simulation to follow the evolution of the galaxy population in clusters \citep{Pillepich2019, Nelson2019b}. This effectively allows us to study the properties of the cluster population down to the faint and ultra-faint dwarf galaxy regime. The scope of this approach is complementary to efforts like SatGen \citep{Jiang2021} and other semi-analytical models \citep[e.g., ][]{Diemer2024,Wan2025}, but similarly aimed. In our case, we wish to preserve as much information as possible from the hydrodynamical  simulation, and therefore we apply the tidal evolution model case-by-case in each cluster and only at resolutions below those captured by the hydrodynamical run. 

This paper is organised as follows. In Section~\ref{sec:data}, we briefly describe the model, simulation and the formalism to use tidal evolution model for subhalos from the simulation. In Section~\ref{sec:results} we present the size-mass relation, satellite mass function, radial distribution and degree of tidal stripping for satellites in Virgo-like clusters, all the way from ultra-faints to massive ellipticals at redshift $z = 0$. We finally conclude with a summary in Section~\ref{sec:conclusions}.

\section{Methods and Satellite Samples}
\label{sec:data}

\subsection{The cosmological simulation -- TNG50}

We use the TNG50-1 cosmological magnetohydrodynamic simulation \citep{Pillepich2019, Nelson2019, Nelson2019b}, the highest resolution run within the IllustrisTNG project. The simulation has a box size of $51.7\,\mathrm{Mpc}$ on a side and includes $2 \times 2160^3$ resolution elements. It uses the {\sc arepo} unstructured moving mesh code \citep{Springel2010} to solve for gravity and magnetohydrodynamics. 

Structures are followed from an initial redshift $z = 127$ until $z=0$.  The mass of a single dark matter particle is $4.5 \times 10^5$~\msun\; and the baryonic mass resolution is $8.5 \times 10^4$~\msun. The softening length for dark matter and stellar particles is $290$ pc at $z = 0$, while it is adaptive with a minimum of $74$ comoving parsecs for gas cells. 

The simulation is based on the concordance $\Lambda$-Cold Dark Matter ($\Lambda$CDM) cosmology with best-fit parameters from \cite{Planck2016}.  The IllustrisTNG project implements a wide range of baryonic processes such as star formation, supernova and AGN feedback, as detailed in \cite{Weinberger2017, Pillepich2018}. Halos were identified using a Friends-of-friends (FoF) algorithm \citep{Davis1985} and subhalos using the {\small SUBFIND} algorithm \citep{Springel2001,Dolag2009}.

We select the three most massive clusters in the TNG50 box  with $M_{200} \geq 10^{13.5}$~\msun. We refer to each of these clusters by their FoF group numbers (FoF-0, FoF-1 and FoF-2). Table~\ref{tab:fof_properties} summarizes the virial quantities for our hosts. Virial quantities are measured at the radius where the enclosed density is $200$ times the critical density of the universe. 

Our systems are in the approximate mass range of objects like the Virgo cluster, with virial mass estimates in the range $(1-9) \times 10^{14}$\msun \citep[see][and references therein]{Kashibadze2020}, Hydra-1 cluster with $M_{200} \sim 1.9 \text{-} 2.1 \times 10^{14}$\msun\; \citep{Girardi1998,Tamura2000} and smaller systems such as the Fornax cluster, with a virial mass estimated at $7 \times 10^{13}$\msun \citep{Drinkwater2001}. 

\begin{table}
\caption{
Properties of three TNG50 galaxy clusters with virial masses above $10^{13.5}$~\msun\, used in this work to investigate the ultra-faint satellite population in galaxy clusters. We refer to each of these galaxy clusters by its FoF group number.
Column (1): Host identifier (FoF group number);
Column (2): Virial mass $M_{200}$ in \msun, defined as the mass enclosed within radius $R_{200}$ where the mean density is 200 times the critical density;
Column (3): Virial radius $R_{200}$ in kpc;
Column (4): Number of {\small SUBFIND} identified satellites with $M_* > 10^{8}$~\msun at $z=0$ within the host in TNG50; Column (5): Total number of subhalos which we track in our modelling.
}
\begin{tabular}{ccccc}
\hline
Host  & $M_{200}$ [\msun]    & $R_{200}$ [kpc] & $N_\mathrm{subs}$ ($M_*>10^8$~\msun) & Tracked $N_\mathrm{subs}$ \\
(1)   & (2)                  & (3)             & (4)                                & (5)                       \\ \hline
FoF-0 & $1.8 \times 10^{14}$ & $1198$          & 389                              & 253814                    \\
FoF-1 & $9.5 \times 10^{13}$ & $959$           & 208                              & 142207                    \\
FoF-2 & $6.5 \times 10^{13}$ & $846$           & 140                              & 95134   \\
\hline            
\end{tabular}
\label{tab:fof_properties}
\end{table}

Throughout this paper, we shall use the term ``central subhalo" or ``central galaxy" to refer to the subhalo (or galaxy) at the center of the gravitational well of each FoF group. All other subhalos/galaxies associated to the system are considered  ``satellites".  A ``surviving" or ``Type-1" satellite is one detected by the {\small SUBFIND} algorithm at $z = 0$ in the FoF halo. In contrast, a ``Type-2" subhalo is one that has been disrupted, and is not found by {\small SUBFIND} at $z = 0$. Note that at least some of these disruptions may be a numerical artifact; our work addresses this issue using the tidal evolution model described in Section~\ref{sec:tidal_model}. 

\subsubsection{Assessing resolution of dark matter subhalos in TNG50}
\label{sec:tng50resolution}

\begin{figure}
    \centering
    \includegraphics[width=\linewidth]{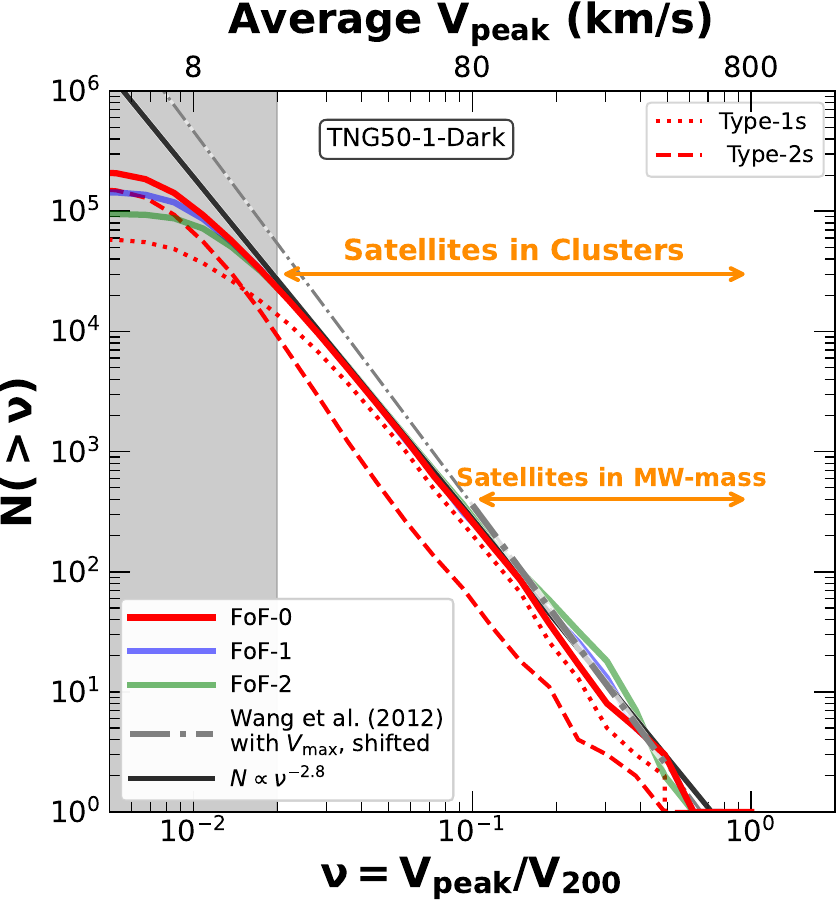}
    \caption{Cumulative subhalo velocity function in the 3 most massive FoF halos in the TNG50-1-Dark simulations. We highlight the contribution of Type-1 and Type-2 satellites as an example for FoF-0 (red curves). The best-fit power-law for the subhalo velocity function is $N \propto \nu^{-2.8}$ (solid black line). The thick gray dash-dotted line illustrates the scaling from \protect\cite{Wang2012} measured in the range $0.1 \le \nu \le 0.5$ (we extrapolate into smaller $\nu$ values using a thinner line). The top axis indicates $V_{\mathrm{peak}}$ when assuming an average virial velocity of our systems $V_{200} = 800$ km/s. Gray shaded region indicates where the subhalo velocity function starts to deviate due to insufficient resolution.  Note that the typical subhalos hosting ultra-faint dwarfs ($V_{\rm peak} \sim 10$ km/s) probe a deeper range of $\nu$ in galaxy clusters than in MW-mass systems (orange arrows). }
    \label{fig:nu_distribution}
\end{figure}

In the absence of baryons, the $\Lambda$CDM scenario makes clear predictions for the number and mass distribution of the satellites of any collapsed host halo, which we refer to as the subhalo/satellite mass function (or, its equivalent, the subhalo velocity function). Given the scale-free nature of the power-spectrum in  $\Lambda$CDM  and of the gravity force, the subhalo mass/velocity function is independent of the host mass, once properly scaled \citep{Kratsov2004,Giocoli2008,Gao2011}. 

More specifically, the number of subhalos above a given velocity threshold, $N_{\rm sub}(>\nu)$, is expected to scale as a power law $N_{\rm sub}(>\nu) \propto \nu^\alpha$ with $\alpha \sim -3$ \citep{Springel2008,BoylanKolchin2010,Wang2012}. Here, $\nu= v_{\rm sat}/V_{200}$ refers to a typical velocity scale of the subhalo compared to the virial velocity of the host. Common assumptions include using the maximum circular velocity $V_{\rm max}$ at $z=0$ or the the maximum of the circular velocity attained over the entire subhalo history, $V_{\rm peak}$, as assumed in our work. 

A flattening or deviation from this power-law behavior at low $\nu$ can therefore be used to determine the regime where numerical limitations become important. We assess the onset of this regime using the {\it dark matter only} version of TNG50 (TNG50-Dark), avoiding complications that are introduced by baryonic effects.

Fig.~\ref{fig:nu_distribution} shows the velocity function of subhalos in terms of $\nu = V_{\mathrm{peak}}/V_{200}$ for our three clusters (thick solid lines). This function includes all subhalos accreted into the cluster; i.e., type-1s and type 2s. 

The subhalo velocity function is well approximated by a power law with slope $-2.8$ in the range $0.02<\nu<0.5$ (solid black line), consistent with previous results \citep[e.g.][]{BoylanKolchin2010}. For illustration, for our most massive system, FoF-0, we divide the contribution of satellites that are still resolved (type-1, thin dotted) from those that are lost in the halo catalog today (type-2, thin dashed line). This decomposition shows that type-1s  dominate the large $\nu$ regimes, while the inclusion of ``orphans" or type-2 satellites is necessary towards smaller $\nu$ values to recover the expected power-law behavior.

Our subhalo velocity function is also in good agreement with that presented in \citet{Wang2012} over the range where the data overlaps (gray thick dot-dashed line). We extrapolate the relation in \citet{Wang2012} toward lower values of $\nu$ using a thinner dot-dashed line. Note that the original relation has been renormalized by a factor of $10$ to account for the fact that we define $\nu$ in terms of $V_{\rm peak}$ and not $V_{\rm max}$ at $z=0$ as \citet{Wang2012} \citep[see also ][]{Santos-Santos2024}. 
We conclude that TNG50 resolves subhalos with $V_{\rm peak} \geq 8-10$ km/s in our host halo mass range, provided ``orphans" or type-2 satellites are taken into account. 

Fig.~\ref{fig:nu_distribution} highlights another important factor: the range in $\nu$ that is necessary to model ultra-faint dwarfs in clusters is much wider than the typical $\nu$ values needed to model ultra-faints in smaller hosts, such as the Milky Way. The orange horizontal arrow shows the range corresponding to $V_{\rm peak}=10$-$50$ km/s assuming a host with $V_{200}=200$ km/s, consistent with a $M_{200} \sim 10^{12}$\msun. Resolving ultra-faint dwarfs in cluster-like environments extends the observational capabilities of testing the self-similarity predicted by $\Lambda$CDM for almost an additional dex in $\nu$ (at fixed luminosity or magnitude detection). 

\subsection{The tidal evolution model}
\label{sec:tidal_model}

\begin{figure}
    \centering
    \includegraphics[width=1\linewidth]{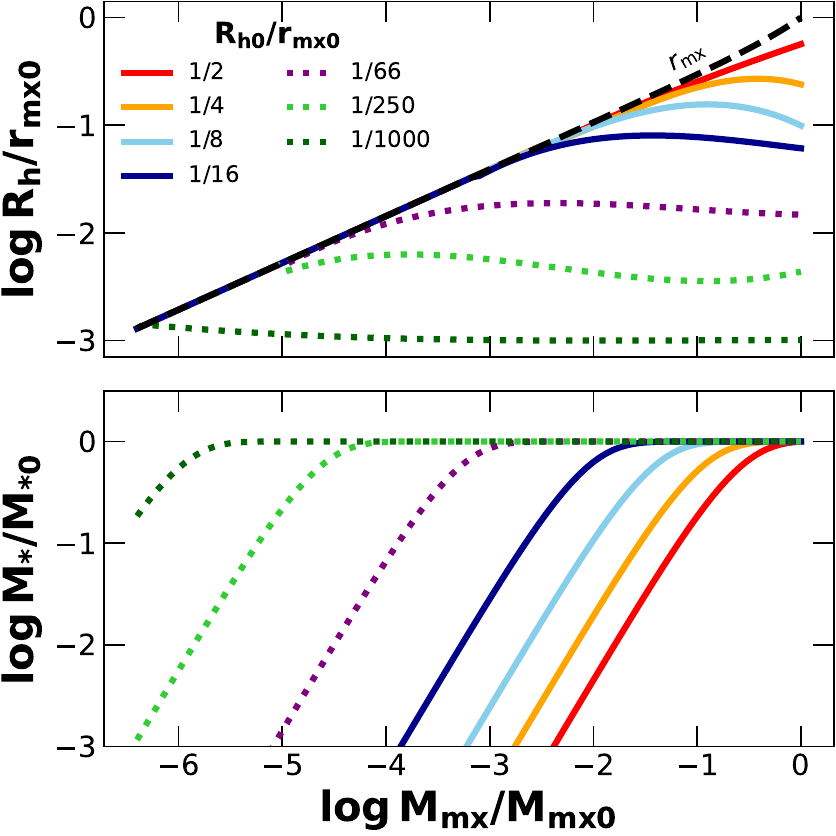}
    \caption{Tidal tracks for the evolution of baryonic properties -- projected (2D) half-light radius $R_\mathrm{h}$ (\textit{upper panel}) and stellar mass $M_{*}$ (\textit{lower panel}) -- of satellite galaxies as a function of amount of stripping measured in terms of the fraction of mass remaining inside $r_\mx$, $M_\mx/M_\mxo$. $M_\mx$ here denotes the mass enclosed within the radius of peak circular velocity $r_\mx$, $M_\mx = r_\mx V_\mx^2 / G$ (while the subscript $0$ refers to the initial values before the tidal evolution). The dashed black line in the upper panel corresponds to the evolution of $r_\mx$ for the dark matter component. The tidal tracks built by extending the empirical model of \protect\citetalias{Errani2022} are shown as colored dotted lines.}
    \label{fig:tidal_track}
\end{figure}

\begin{figure*}
    \centering
    \includegraphics[width = 1\linewidth]{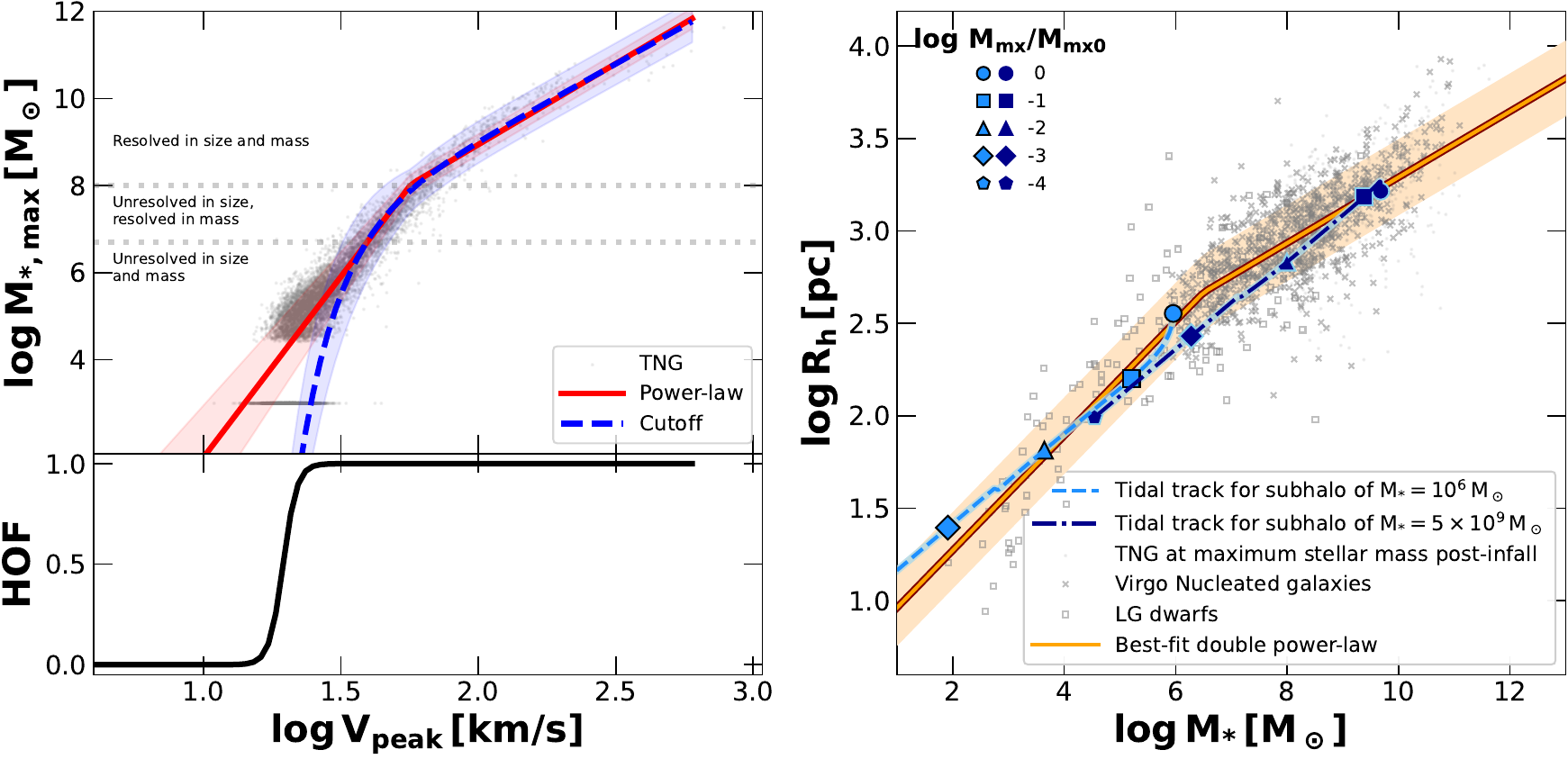}
    \caption{\textit{Top Left}: The abundance matching relation for TNG50 simulation along with two chosen extrapolations adopted for the initial stellar mass - halo mass relation. Grey dots show maximum stellar mass $(M_\mathrm{*, max})$ vs peak maximum circular velocity $(V_\mathrm{peak})$ for Type-1 satellites from FoFs-0, 1 and 2 in TNG50. Subhalos without stellar particles in the simulation are artificially shown at $M_\mathrm{*}=10^3$\msun. Horizontal dotted gray lines show our thresholds to consider an object resolved/unresolved. The ``power-law'' and ``cutoff'' models used to populate the unresolved subhalos are shown as red solid and blue dashed lines respectively.\\
    \textit{Bottom Left}: The halo occupation fraction (HOF) is shown as a black solid line. 
    \textit{Right}: To assign sizes at infall to subhalos unresolved in size, we fit a double power-law (shown in orange) to observed dwarf galaxies in Virgo (gray crosses) and ultra-faint dwarfs in the Local Group (gray open squares). Grey dots represent resolved TNG satellites from FoFs-0, 1, and 2 at the time of $M_{*,\rm max}$. We also show the tidal evolution of two selected satellites with initial stellar masses of $\approx 10^6$ (light blue, $R_h/r_\mxo = 0.074$) and $5 \times 10^9$\msun (dark blue, $R_h/r_\mxo = 0.075$). Markers along the tidal tracks indicate different fractions of $M_\mathrm{mx} / M_\mathrm{mx0}$. Notably, the tidal evolution closely follows the best-fit double power-law to the data.}
    \label{fig:abundance}
\end{figure*}

\begin{figure}
    \centering
    \includegraphics[width = 1\linewidth]{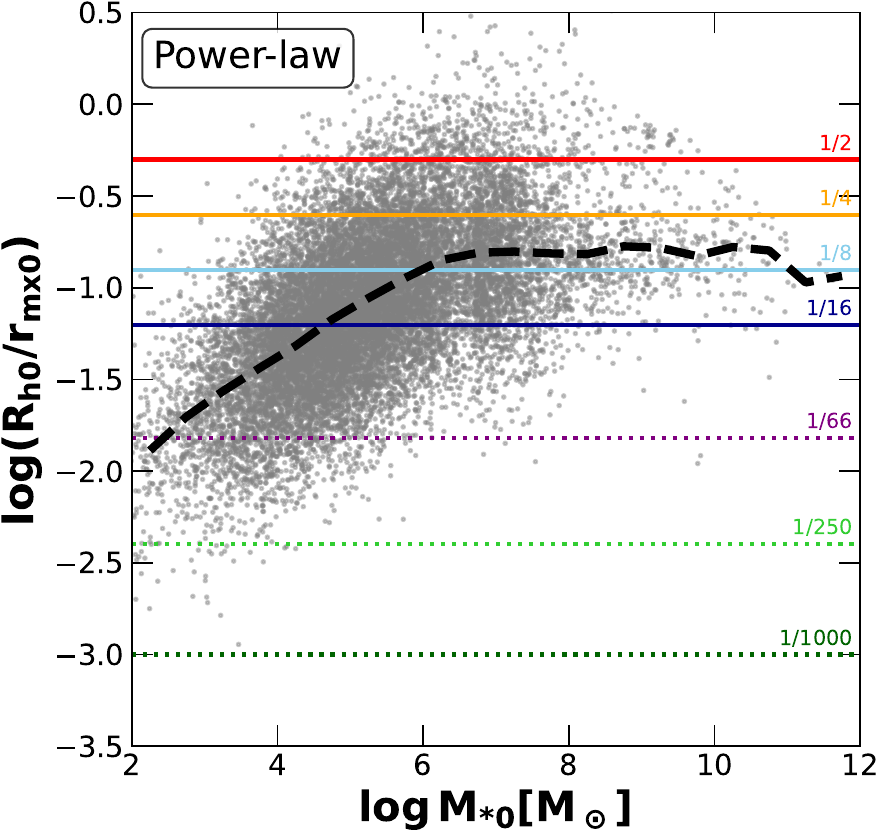}
    \caption{The initial stellar segregation, defined as the ratio between the projected half light radius and $r_\mathrm{mx}$. for the stars in their dark matter subhalos as a function of the infall stellar mass $(M_{*0})$. Here, unresolved subhalos are populated using the power-law extrapolation. $R_\mathrm{h0}^\mathrm{pl}$ is the projected infall stellar half-light radius, whereas $r_\mathrm{mx0}$ is the radius at which the circular velocity profile of the subhalo peaks at infall. The black dashed line indicates the median $R_{h0} / r_\mxo$ in bins of stellar mass having a width of 0.5 dex. The horizontal lines indicate the different initial segregations for which stellar tidal tracks are available from \protect\citetalias{Errani2022}. For each satellite galaxy, we interpolate between the tidal track that best-approximates the initial segregation of its stellar component $R_{h0} / r_\mxo$.}
    \label{fig:segregations}
\end{figure}

The tidal evolution of a cold dark matter subhalo orbiting in the gravitational potential of a more massive host can be parametrized in terms of a few simple quantities \citep{Hayashi2003, Penarrubia2008,Errani2015}. Subhalos are modelled to follow an NFW profile at accretion, and are subsequently trucated by tides. We characterize the subhalo strucure in terms of the radius at which circular velocity curve peaks ($r_\mathrm{mx}$) and the maximum circular velocity it reaches ($V_\mathrm{mx}$), or, alternatively of its mass inside $r_\mathrm{mx}$ ($M_\mathrm{mx}$). We use the model in \citet[][\citetalias{Errani2021} hereafter]{Errani2021}, who show that NFW halos evolve tidally following a "tidal track" given by 
\begin{equation}
    V_\mathrm{mx}/V_\mathrm{mx0} = 2^\alpha (r_\mathrm{mx}/r_\mathrm{mx0})^\beta [1 + (r_\mathrm{mx}/r_\mathrm{mx0})^2]^{-\alpha}
    \label{eq:tidal_track}
\end{equation}
\noindent
with $\alpha = 0.4$ and $\beta = 0.65$. Subscript 0 indicates initial values throughout this paper. Tidal tracks are independent of orbital eccentricity or time elapsed, and is solely dependent on the total amount of mass lost since infall (parametrized by $M_{\rm mx}/M_{\rm mx0}$, see Section 3.6 of \citetalias{Errani2021}).

Under the assumption of a dark matter dominated system, which is particularly appropriate for faint and ultra-faint dwarfs, the evolution of a stellar component under tidal disruption can be added to this model. The evolution depend mainly on the radial segregation between stars and dark matter, which  is parametrized by the ratio between the initial half-mass radius of the stars and the characteristic radius of the halo, $R_\mathrm{h0}/r_\mathrm{mx0}$ \citet[][\citetalias{Errani2022} hereafter]{Errani2022}. With this assumption, it is possible to compute tidal tracks that describe the evolution of the remnant stellar mass (or luminosity), $M_*$, remnant stellar projected (2D) half-mass radius, $R_\mathrm{h}$, and velocity dispersion $\sigma$. 
In particular, we choose here the tidal tracks that correspond to a 3D exponential light profile.

This is  shown in Fig.~\ref{fig:tidal_track}, which illustrates the evolution of the stellar half-mass radius ($R_\mathrm{h}$) and the total stellar mass ($M_*$) assuming different initial segregations (different colors as labelled). Curves for $R_\mathrm{h0}/r_\mathrm{mx0} = 1/2, 1/4, 1/8$ and $1/16$ were originally included in \citetalias{Errani2022}. 
We have also generated  tidal tracks for $R_\mathrm{h0}/r_\mathrm{mx0} = 1/66, 1/250$ and $1/1000$ (dotted lines), which are necessary to model our fainter galaxies. In the top panel, we also include the evolution of the size ($r_\mathrm{mx}/r_\mathrm{mx0}$) for the NFW dark matter component (black curve). 

Note that the size of the stellar component is hardly affected at the beginning, but, once the halo has been stripped to radii comparable to that of the stars, the evolution of $R_h$ closely follows that of $r_\mathrm{mx}$. On the other hand, the stellar mass or luminosity (bottom panel of Fig.~\ref{fig:tidal_track}) seems more sensitive to stripping. 

In practice, the basic ingredients required to model the properties of the satellite galaxy at any given time are: (i) Orbital properties: infall time, orbital time $T_\mathrm{orb}$, pericentric ($r_\mathrm{peri}$) and apocentric ($r_\mathrm{apo}$) radii. (ii) Initial dark matter properties $r_\mxo, V_\mxo$ and $M_\mxo$; (iii) Initial stellar properties -- $R_\mathrm{h0}$ and $M_{*0}$. We describe them individually in what follows.

\subsection{All ingredients together: modeling infalling subhalos in TNG50}
\label{ssec:input_extraction}

We have used the merger trees to compile a catalogue of all subhalos that have entered the FoF group of our 3 hosts throughout their history. Table~\ref{tab:fof_properties} summarizes the numbers of subhalos of each type analyzed per host. The infall snapshot for each subhalo was determined by examining the main progenitor branch of the subhalo back on time to identify the first snapshot when the subhalo of interest first joins the host FoF halo.

For simplicity, in the tidal evolution model we assume that the mass profile of our 3 cluster hosts, may be approximated by an isothermal sphere, with circular velocity, $V_0$, chosen to match the circular velocity at the virial radius at $z = 0$. This typically provides a good description of the mass distribution in the host including all (dark matter and baryonic) components. 

We have explicitly checked that the lack of time evolution in the host assumed here does not fundamentally change the statistical properties of our samples. For example, to take into account the growth of the host potential with time, we have run a version of our model in which we used the $V_0$ that best-fit the host profile at infall for each satellite, instead of that at $z=0$. We found no qualitative difference in the results, except for a small subset of early infalling halos for which this change results in appreciable differences in the predicted bound mass (for instance, fewer than $8\%$ of the satellites show changes in their predicted stellar mass larger than a factor $\times 3$ when changing the $V_0$ assumption).  

\begin{figure*}
    \centering
    \includegraphics[width = 1\linewidth]{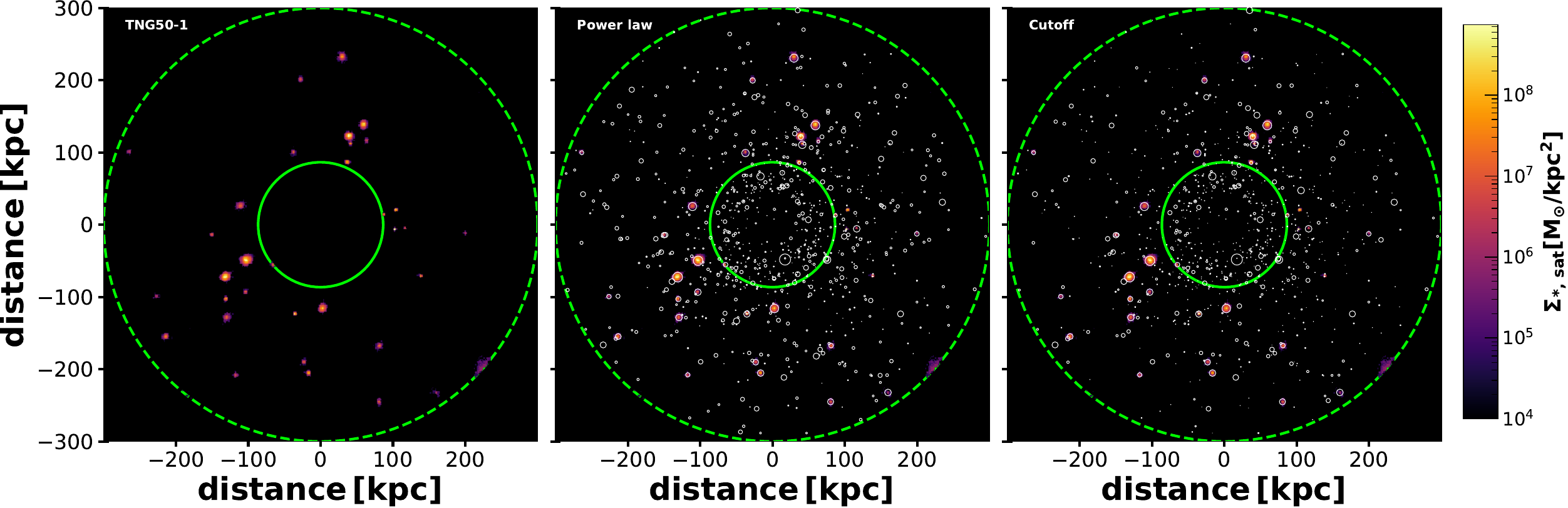}
    \caption{Projected thin slice of FoF-0 from TNG50-1 in the xy-plane, excluding the central subhalo for the simulation versus power-law and cutoff models. The region shown corresponds to $\sqrt{x^2 + y^2} < 300$ kpc and $|z| < 50$ kpc in cluster-centric coordinates. This region is encircled by a green dashed circle. \textit{Left panel}: Stellar particles at $z = 0$ are binned to represent the surface brightness of the satellite galaxies. The stellar particles from the central subhalo have been excluded. \textit{Central and right panels}: Extensions for the predictions of the TNG50 simulation based on assumed power-law (center) and cutoff (right) abundance matching relations. The green solid circle indicates the stellar half-mass radius for the central subhalo. 
    Satellite galaxies are depicted as white circles with a radius scaled to 10 times the projected half-light radius $R_h$. Satellite in the inner 50 kpc have been excluded since they would remain undetectable near bright central galaxy. }
    \label{fig:slice}
\end{figure*}

One of the main goals of our modeling is to correct for the luminous satellites that are missing from the TNG50 simulation at $z=0$ because either (i) the expected stellar mass/luminosity is below the resolution limit, or (ii) the subhalo gets fully (artificially) disrupted in the tidal field of the host. 

To address the first item, we assign a mass  to the stellar component of all under-resolved subhalos at infall. The stellar mass depends mainly on the peak subhalo velocity (in practice, its maximum circular velocity at infall). We report results for two different assumptions, a "power-law model" and a "cutoff model", as in \citet{Santos-Santos2022} calibrated to the resolved galaxies in TNG50. A size is also assigned to the stellar component in the case that is considered unresolved, using a double power-law motivated by the observed sizes of dwarf galaxies in the Local Group. Details of the modeling are shown in Fig.~\ref{fig:abundance}. 

Note that the chosen assignments only modify the stellar mass and sizes of low-mass subhalos. Satellites that are considered resolved and survive to $z=0$ in the simulation are not affected by this modeling. Only for $M_{*,\rm max} < 5 \times 10^{6}$\msun\; we consider the stellar mass ``unresolved". The threshold to consider an object unresolved {\it in size} is  $M_* < 10^8$\msun, based on the scales where spurious heating becomes significant following \citet{Ludlow2023}.  

To account for the effects of reionization suppressing star formation in low mass halos, we further adopt a Halo Occupation Fraction (HOF) that quantifies the fraction of dark matter halos that may host galaxies within a given mass range \citep[e.g., see ][]{Benitez-Llambay2020}. In this work, we adopt the analytical fit for the HOF from \cite{Santos-Santos2024} to determine the probability that a dark matter halo hosts a galaxy, which is based on results from GALFORM model for galaxy formation \citep{Lacey2016}. 

Hereafter, the terms ``power-law'' and ``cutoff'' assumptions refer to unresolved subhalos having been populated using these extrapolations respectively, before evolving the subhalos using the tidal evolution model. Quantities computed with these assumptions are indicated by superscripts ``pl'' and ``co'' respectively.

\begin{figure*}
    \centering
    \includegraphics[width=1\linewidth]{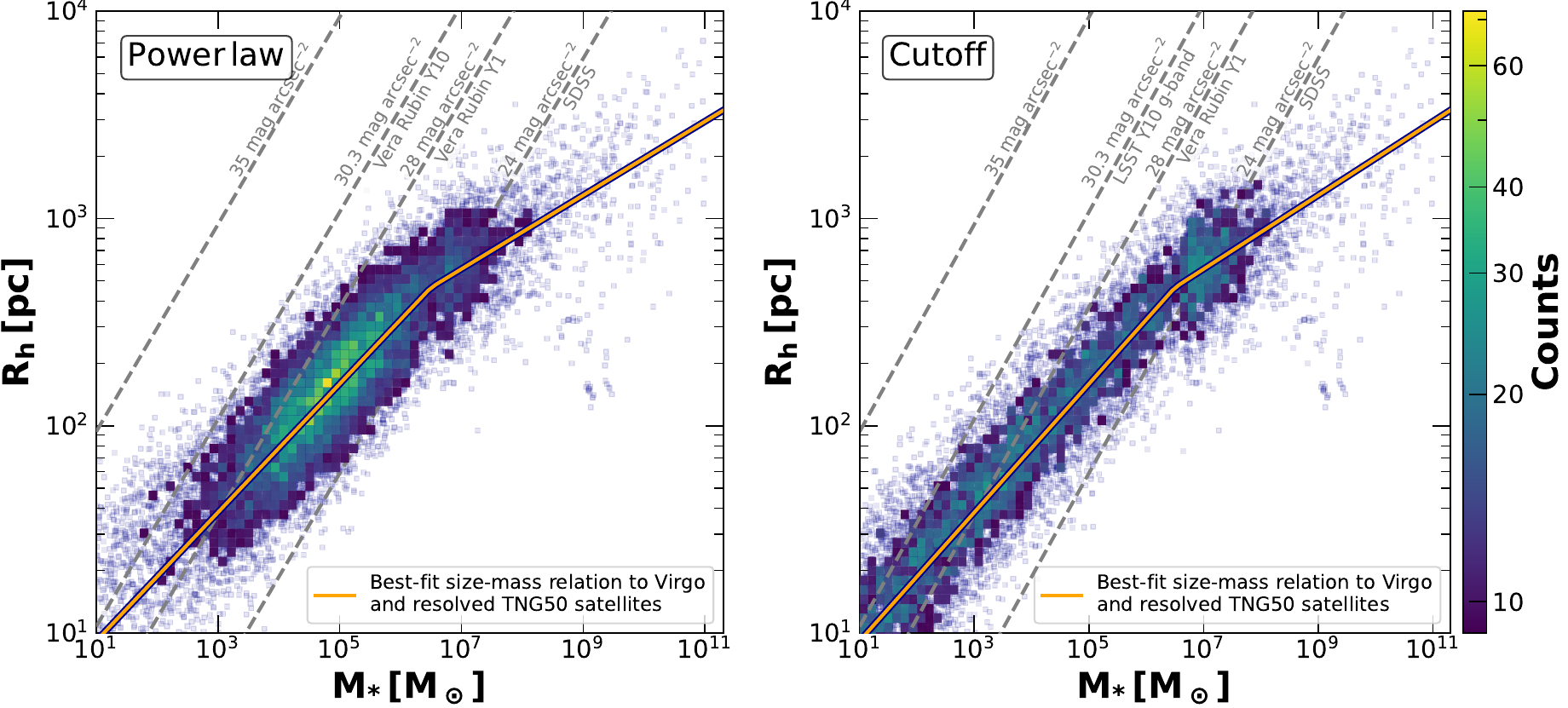}
	\caption{2D histogram for the combined size-mass relation for subhalos from all three FoF halos in TNG. The left and right panels correspond to the power-law and cutoff models, respectively.  The dashed grey lines show lines of constant surface brightness. Yellow solid line shows the best-fit double power-law to observations of the galaxies belonging to Local Group and Virgo clusters. Most of the subhalos evolve along the initial relation assumed.}
    \label{fig:size_mass}
\end{figure*}

\subsection{An illustrative example of the model}

With all the necessary infall properties at hand, we calculate the initial segregation $(R_\mathrm{h0}/r_\mxo)$ for each TNG50 subhalo by taking the ratio of the initial 2D half-mass radius ($R_\mathrm{h,0}$) and $r_\mxo$. Fig.~\ref{fig:segregations} shows the initial segregations for TNG50 subhalos in all three FoF halos considered, for the case of the power-law abundance matching case. Dwarf galaxies become intrinsically more embedded within their dark matter subhalos as stellar mass decreases, i.e, in our model, the ratio $R_{h0}/r_\mxo$ decreases with decreasing stellar mass $M_*$. For example, a galaxy of $M_* = 5 \times 10^9$ \msun, we typically find $R_{h0}/r_\mxo \sim 0.15$, whereas for $M_* = 10^4$ \msun, typically we fins $R_{h0}/r_\mxo \sim 0.03$. The stellar tidal tracks have been pre-computed for initial segregations of $R_\mathrm{h0}/r_\mxo = \{1/2, 1/4, 1/8, 1/16, 1/66, 1/250, 1/1000\}$, as shown in Figure~\ref{fig:tidal_track}. For each satellite, we then interpolate linearly between the results from two precomputed tidal tracks that bracket their initial stellar segregation. For subhalos with initial segregation outside of these precomputed ranges —-i.e., $R_\mathrm{h0}/r_\mathrm{mx} > 1/2$ or $< 1/1000$-— we assign them to the tidal track for $1/2$ or $1/1000$, respectively. 

We calculate the cluster-centric positions of satellites at $z=0$ as follows. For each Type-1 subhalo, we directly obtained the position of the subhalo (using the definition of the potential minimum, or ``SubhaloPOS'') from the group catalogue and use the position of the central subhalo in the FoF as cluster-center. For Type-2 subhalos (those considered merged or disrupted in the simulation), we tracked the most bound particle (MBP) of the subhalo from its last two snapshots identified in the merger tree (i.e., two snapshots before the subhalo is lost in Subfind catalogs) and consider the average of their coordinates as final position of the satellite.

 As an illustration of the capabilities of our method, we show in Fig.~\ref{fig:slice} a thin $300\,\mathrm{kpc}$ radius slice (50 kpc depth) centered on our largest FoF group. Since the central galaxy would outshine any satellite within 50 kpc, we remove any satellite predicted to lie within a 3D sphere 50 kpc from the central galaxy. The left panel includes all satellites considered resolved in TNG50 (i.e., which appear in the Subfind catalog and have $M_*> 5 \times 10^{6}$\msun). The results of applying our tidal evolution method are displayed over the same background image in the middle and right panels for the power-law and cut-off models, respectively. White circles indicate the positions of surviving satellites in our catalogs, with sizes proportional to the projected half-light radius calculated for each object at $z=0$. Our method is able to follow the original surviving satellites in TNG50 and, in addition, keeps track and predict properties for subhalos that are unresolved or totally disrupted in the original cosmological simulation. Although the total number of observable satellites at $z = 0$ is sensitive to the adopted choice of power-law or cutoff model (see Fig.~\ref{fig:abundance}), we find that the radial profile of the satellite number density is largely insensitive to this choice. We explore these results in detail below.

Hereafter, quantities with no subscript correspond to their values at $z=0$, unless otherwise stated. On the other hand, a subscript of ``0'' indicates values used at infall for the model before evolution, as mentioned before. 

\section{Results}
\label{sec:results}

\subsection{The stellar mass - size relation}
\label{ssec:mass_size}

Fig.~\ref{fig:size_mass} shows the $M_*$-size relation at $z=0$ for galaxies in  the ``power-law'' and ``cutoff'' models. The evolved relation stays close to the infall one, whose best-fit is shown by the orange line in each panel. This is mostly because tidal stripping moves objects along the relation (see example tracks in the right panel of Fig.~\ref{fig:abundance}), with little impact on the scatter. The r.m.s (measured as dispersion in size at fixed $M_*$) of the initial infall distribution is $0.20$ dex, while the evolved one shown here at $z=0$ is $0.22$ for both, the power-law and cut-off models. This is a feature of tidal evolution models under the assumption of an exponential (3D) light distribution \citep[e.g., ][]{Errani2024b}.

Gray dashed lines in Fig.~\ref{fig:size_mass} indicate surface brightness limits $\Sigma = 24, 28, 30.3$ and $35$ mag/arcsec$^2$ (calculated as $\Sigma = L/(\pi  R_h^2)$ assuming a mass-to-light ratio unity, $M_*/L=1~\mathrm{M_\odot/L_\odot}$). These may be compared with typical sensitivity limits $\sim 24$ mag/arcsec$^2$ for SDSS in the $r$-band \cite{Du2015, Yi2022} or $28$-$30$ mag/arcsec$^2$ for Vera-Rubin year-one to year-10 estimates \cite{Brough2020}. The $35$ mag/arcsec$^2$ limit shows the surface brightness sensitivity needed to include all satellites predicted by our model. 

Since tidal stripping moves satellites roughly along the mass-size relation, we expect the upcoming discovery of ultra-faint dwarfs in groups and clusters with Vera Rubin and other high sensitivity surveys to follow a size distribution similar to the ultra-faint population in the Local Group. 

Figure~\ref{fig:size_mass} highlights an important difference between the two stellar mass-halo mass models: the cutoff model (right) predicts a greater number of ultra-faint galaxies with $M_* \lesssim 10^3$ \msun\; than the power-law model (left panel). This difference arises as a consequence of the halo occupation fraction assumed, which heavily limits the fraction of low mass halos with $V_{\rm peak}<20$ km/s, corresponding to $M_* < 10^4$\msun, that can host a luminous counterpart in the power-law model (see left panel in Fig.~\ref{fig:abundance}). The halo occupation fraction plays a smaller role in the cutoff model, since most ``luminous" objects with $M_*> 100$\msun\; correspond to halos with $V_{\rm peak}>20$ km/s, where the occupation fraction climbs to $\sim 1$. 

\subsection{Satellite mass function}
\begin{figure*}
    \centering
    \includegraphics[width = 1 \linewidth]{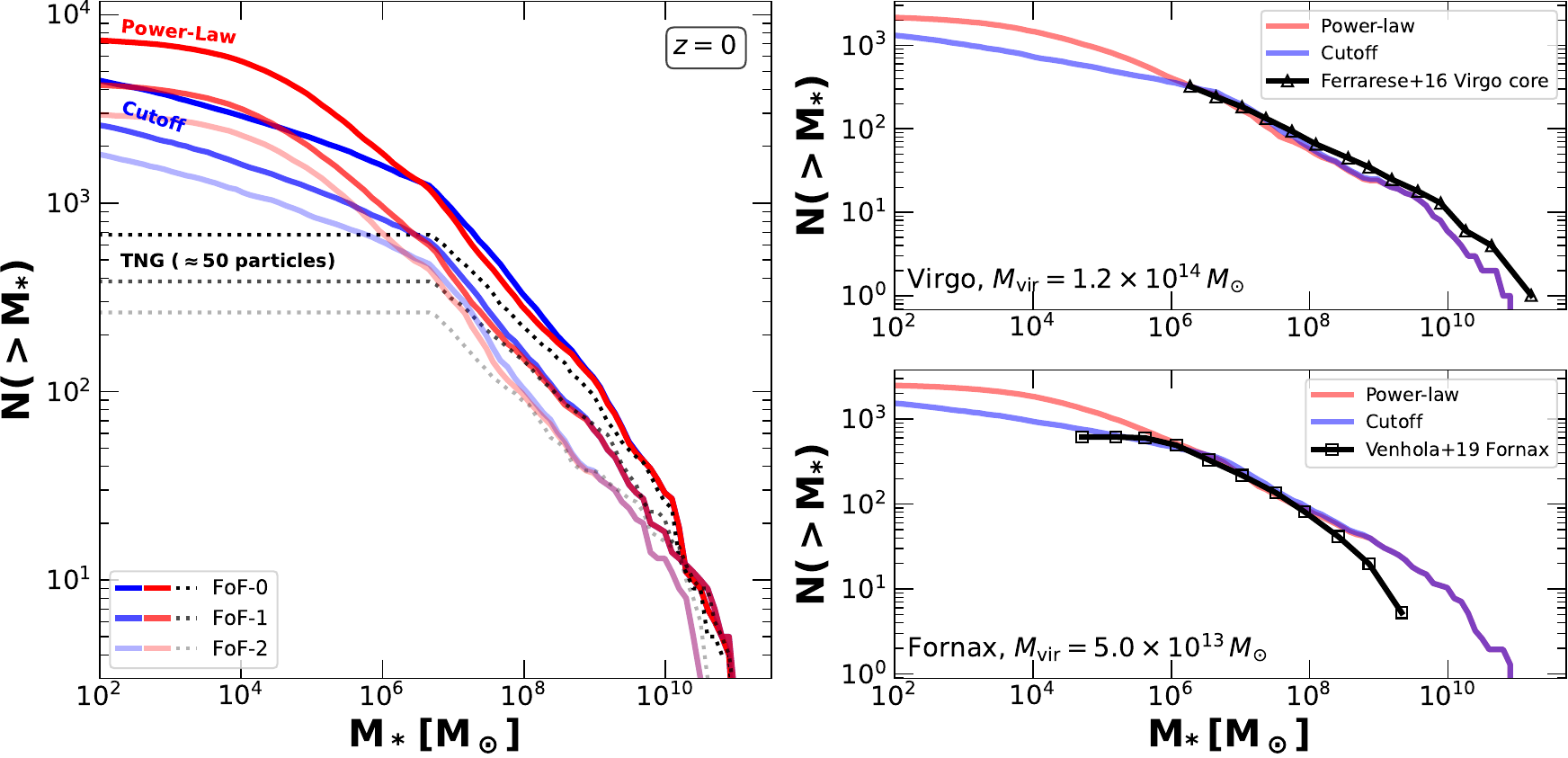}
	\caption{\textit{Left}:The projected satellite mass function for all three FoF halos. Power-law models are shown in red, and cutoff models are shown in blue. Dotted lines represent the satellite mass functions for the resolved Type-1 satellites in the TNG50 simulation. Different transparencies represent each FoF halo. \textit{Right}: projected median satellite stellar mass function for the three FoF groups considered within 309 kpc (top) and 700 kpc (bottom) in red/blue for the power-law and cutoff models. Normalization of this average mass function to observational data in the literature for the Virgo cluster \protect\cite{Ferrarese2016} (upper panel, black solid line with triangular markers) and Fornax \citep{Venhola2019} (lower panel, black solid line with square markers) suggest virial masses for each system as quoted in the bottom left of each panel.}
    \label{fig:smf_3panel}
\end{figure*}

\begin{figure}
    \centering
    \includegraphics[width = 1 \linewidth]{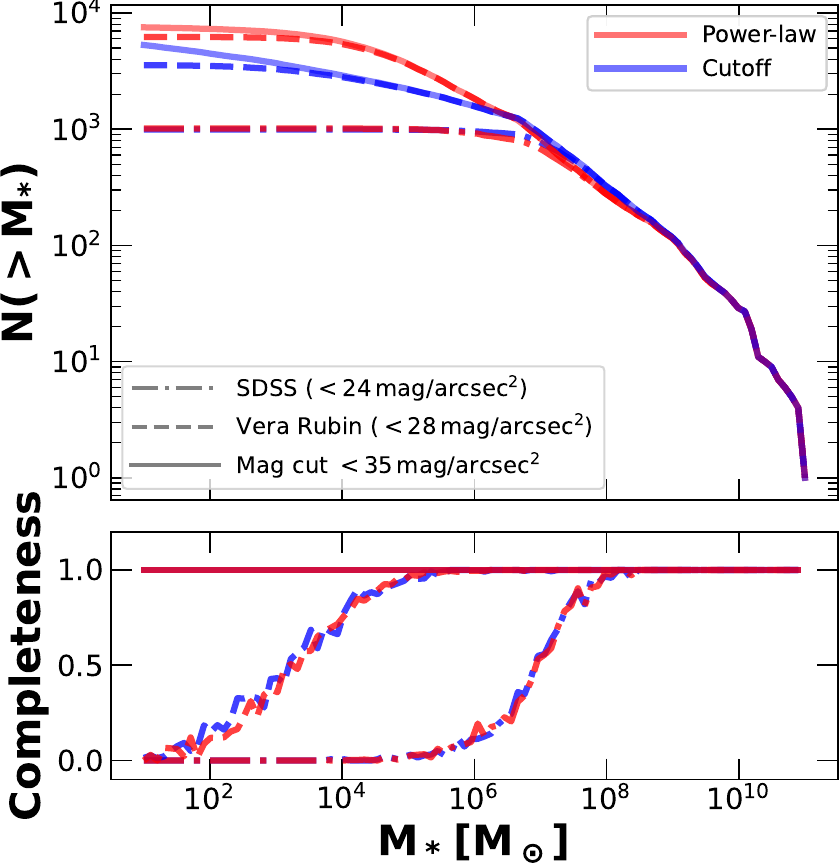}
    \caption{\textit{Upper panel}: predictions for the satellite mass function in FoF-0 with surface brightness cuts at 24, 28 and 35 $\mathrm{mag/arcsec^2}$ (dash-dotted, dashed and solid lines, respectively). \textit{Lower panel}: Completeness, defined as the fraction of satellites at a given $M_*$ that satisfies the surface brightness cut. Current surveys such as the SDSS can probe up to $24 \, \mathrm{mag/arcsec^2}$, and are thus complete down to a stellar mass of about $10^8$\msun. However, Vera Rubin could reach a depths of $28 \rm - 30\,\mathrm{mag/arcsec^2}$, and expand completness down to a stellar mass $\sim 10^5$\msun. A surface brightness cut of 35 $\mathrm{mag/arcsec^2}$ shows a completeness of unity for the entire range.}
    \label{fig:completeness}
\end{figure}

We compare next the satellite stellar mass function predicted by our model with the results for surviving satellites in TNG50. 
The left panel of Fig.~\ref{fig:smf_3panel} shows the cumulative function for our three TNG50 clusters. We consider all satellites within the (3D) virial radius of their hosts. Different line intensity identifies each of the three groups FoF groups 0-2, with the faintest lines corresponding to our lowest mass system (FoF-2). Black dotted lines show the results taken directly from the halo catalogs of the TNG50 simulation, while solid red/blue is used for our model in the power-law or cutoff model, respectively. 

Our model predicts a steep increase in the number of satellites with decreasing mass, which starts to deviate from the simulation at stellar masses $M_* \sim 10^8$\msun. Most importantly, our method allows us to predict the satellite population even below the $\sim 50$ stellar particle limit in TNG50, or $M_*=5 \times 10^6$\msun, seen here by the pivot point where the dotted line becomes horizontal.

Our model points to a very large population of faint and ultra-faint dwarfs in groups and clusters,  with 2000 - 7000 satellites with $M_* > 100$\msun\; in $M_{200} \sim 10^{14}$\msun\; systems.  In general, the power-law model results in a more numerous satellite population, $\sim \times 1.5$ larger than  that of the cut-off model at $M_* = 10^2$ \msun. This is because the power-law relation is able to populate lower mass subhalos, which are more abundant than those populated by galaxies in the cut-off model.

The difference in the slope of the cutoff and the power-law models has also a sizeable impact on the  low-mass end slope of the satellite mass function, offering an avenue to observationally constrain the halo mass - stellar mass relation in the future.  Interestingly, predictions for the power-law model show a flattening for $M_*<10^4$\msun, which in our model are due to the effects of reionization -- parametrized here through the halo occupation fraction. This reduces considerably the likelihood of small mass halos to host faint dwarfs below that stellar mass, creating a stagnation in the number of ultra-faints predicted (this was also visible in the stellar mass - size relation, discussed in Sec.~\ref{ssec:mass_size}). 

Numerical effects associated to artificial disruption may strongly impact these predictions, even in systems considered marginally  ``resolved" in the parent simulation. For instance, considering satellites with $M_* \geq 10^7$\msun\; (equivalent to $\sim 100$ stellar particles and a median $\sim 3800$ number of dark matter particles at $z = 0$ in TNG50-1), we find  48\%  (power-law) and 62\% (cut-off) increase in the number of satellites predicted within $r_{200}$ in our model compared to the halo catalogs in TNG50. This depletion increases with decreasing satellite masses and in general is stronger in the inner regions of the hosts, in qualitatively agreement with previous results \citep[e.g., ][]{Green2021, Benson2022, Diemer2024}. 

\subsection{Satellite mass functions as virial mass estimators}

The satellite mass functions in the left panel of Fig.~\ref{fig:smf_3panel} naturally vary vertically from each other scaling by their virial masses: more massive host halos host proportionally larger number of satellites. A similar effect has been reported previously for the dark matter mass subhalo mass function \citep[e.g., ][]{Wang2012}. We can therefore use these predictions in combination with observational constraints on the satellite stellar mass function to put constraints on the virial mass of known galaxy clusters where such data is available. 

As an example, we apply this idea to the Virgo cluster, shown on the top right panel of Fig.~\ref{fig:smf_3panel}, where we use data from \citet{Ferrarese2016} on the stellar mass function within its core radius ($309 \, \mathrm{kpc}$). Similarly, on the bottom right panel we take results from \citet{Venhola2019} for the satellite mass function of the Fornax cluster within its virial radius ($700 \, \mathrm{kpc}$). 

To calculate the virial masses of these clusters, we project the satellite distributions of the three FoF groups in our sample in the  XY, YZ ans ZX planes and compute the corresponding satellite stellar mass functions counting satellites within each FoF group that are in projection within either $300$ kpc (for the Virgo case) or within the virial radius (to compare to the data of Fornax). We find the best-fit slope and intercept for a relation linking the total number of satellites with $M_*>10^7$\msun\; within these radii and the virial radius of the hosts.  We find $n_{\rm sat}(M_*>10^7~\mathrm{M_\odot}) = N M_{200}^\alpha$, with $N=10^{-6.21}$ and $\alpha =0.60$ when considering the satellites inside the core $300$ kpc regions of the clusters, and $N = 10^{-7.76}$ and $\alpha = 0.74$ when we consider the satellites inside virial radius.

With this calibrated relation, we use the observed number of satellites to estimate virial masses for Virgo and Fornax, obtaining $M_{200} = 1.2 \times 10^{14}$\msun\; and $5.0 \times 10^{13}$\msun, respectively. The right panels of Fig.~\ref{fig:smf_3panel} show the average satellite mass function in our model normalized vertically as described above for the power-law and cut-off models (red and blue solid lines, respectively). 

Encouragingly, the shapes of the observed satellite mass functions seem in reasonable agreement with our model, in particular for Virgo where the data for all galaxies above a threshold $M_* \approx 10^6$ \msun  is available. Note that in the case of Fornax,  \citet{Venhola2019} reports information only for $M_* < 2 \times 10^9$\msun. The virial mass estimated with this method are in good agreement with previous estimates based on different tracers, highlighting the potential of satellite mass functions in the future to estimate virial masses of host groups and clusters. 

\subsection{Completeness limits}

The detectability of satellites in a given optical and spectroscopical survey will be ultimately given not only by their stellar mass, but also by their surface brightness, with more diffuse systems being more prone to be missed in catalogs. We use the predictions of mass and size in our model to investigate the impact of different surface brightness cuts in the measured abundance of satellites. 

We focus on our largest mass FoF system (FoF-0), and calculate the satellite stellar mass function in Fig.~\ref{fig:completeness} assuming surface brightness cuts: $24$, $28$ and $35$ mag/arcsec$^2$. Here we assume for simplicity a mass-to-light ratio unity and surface brightness corresponds to that average within the effective radii. For the brightest cut, $24$ mag/arcsec$^2$, comparable to present day large galaxy surveys like SDSS, we can achieve 90\% completness at $M_* \sim 10^8$\msun. For the next generation surveys like those from Vera Rubin observatory, assuming surface brightness limit  $28$ mag/arcsec$^2$ we expect satellite systems to be constrained down to $M_* \sim 10^5$\msun\; assuming similar completness levels, enabling the study of classical dwarf - like systems outside the Milky Way and beginning to probe the ultra-faint regime (75\% completness for $M_* \sim 10^5$\msun\; and 65\% for $M_* \sim 10^4$\msun). Sensitivities in the order of $32$-$35$ mag/arcsec$^2$ are necessary to fully explore ultra-faints with $M_* \sim 100$\msun\; and above.  

\subsection{Radial distribution of satellites}

\begin{figure*}
    \centering
    \includegraphics[width = 1 \linewidth]{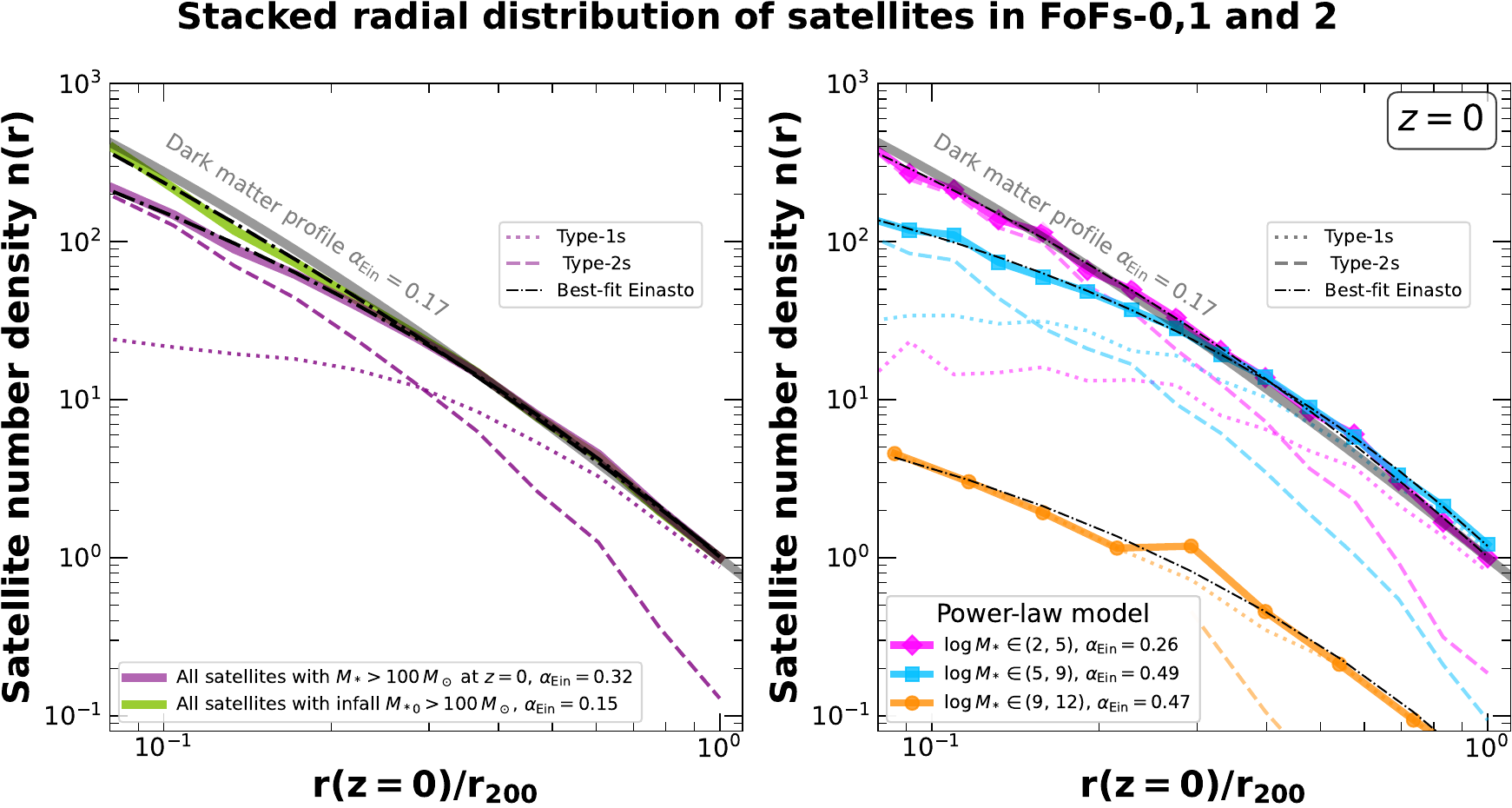}
	\caption{Left: predicted radial distribution of satellites with $M_*>100$\msun\; at $z=0$ (purple solid) compared to the dark matter distribution in the hosts (gray solid line). We have stacked all 3 clusters using the power-law version of the model. Satellites show a more cored distribution than the underlying dark matter particles. This bias reduces when considering all satellites that ever infall in the clusters, even if their present-day stellar mass is below $M_*=100$\msun\; (light green solid). We quote $\alpha_\mathrm{Ein}$ values corresponding to Einasto fits to individual samples in the labels. Note that for all satellites with $M_*>100$\msun\; today, the predicted satellite distribution when considering only Type-1 satellites (dashed brown) is severely suppressed in the inner regions compared to the final result when Type-2 objects are included (dotted brown). Right panel: we explore the stellar mass dependence on the radial distribution of satellites predicted for galaxies with different stellar mass cuts (at $z=0$): massive ($M_*>10^9$\msun, orange), classic dwarfs ($M_*=[10^5 \rm - 10^9]$\msun, light blue) and ultra-faints ($M_*=[10^2 \rm - 10^5]$\msun, magenta). Cuts in $M_*$ result on increasingly biased satellite distributions with respect to the dark matter in the host due to preferential stripping in the inner regions. Ultra-faint dwarfs are therefore better tracers of the dark matter than the more luminous populations.}
    \label{fig:radial_dist}
\end{figure*}

We show the number radial density profile of satellites, $n(r)$, in Fig.~\ref{fig:radial_dist} in units of their cluster-centric distance at $z=0$, normalized to the virial radius of each host system. Results displayed correspond to the power-law  model, but are qualitatively similar for the cut-off model. The left panel in Fig.~\ref{fig:radial_dist} indicates that considering all satellites with $M_*>100$\msun\; today (brown solid line) results in a radial distribution that is shallower than the dark matter in the host potential (gray). We can quantify this in terms of the Einasto profile \citep{Einasto1965,Navarro2004}:
%
\begin{equation}
    \ln (n(r)/n_{-2}) = -(2/\alpha_\mathrm{Ein})[(r/r_{-2})^{-\alpha_\mathrm{Ein}} - 1],
\end{equation}
%
\noindent
where $n_{-2}, r_{-2}$ and $\alpha_\mathrm{Ein}$ are free parameters. Einasto profiles have been shown to provide a good description for the cuspy dark matter distribution in dark matter halos within $\Lambda$CDM with $\alpha_\mathrm{Ein} \approx 0.2$ closely resembling a typical NFW profile \citep{Navarro2010}. Cored dark matter profiles result in a larger value of $\alpha_\mathrm{Ein}$ than the NFW-equivalent values of $\approx 0.2$. 

We find that all satellites with $M_*>100$\msun\; within the virial radius of the host halos are best-fit with an Einasto profile with $\alpha_\mathrm{Ein}=0.32$, and therefore shallower than an NFW expected for the dark matter. In fact, the dark matter density profile in the host clusters is best represented by an Einasto fit with $\alpha_\mathrm{Ein}=0.17$, clearly more centrally concentrated than the present day satellites.

Note that we divide the present-day satellite distribution in contribution from the ``Type-1" and ``Type-2" satellites (dashed and dotted lines, respectively). The inclusion of Type-2s is crucial to model the satellite population in the inner regions of our systems, while Type-1 objects dominate the satellite population outside $r/r_{200} \sim 0.4$. The significantly shallower radial distribution of Type-1 satellites compared to the dark matter in the host is in agreement with previous claims in the literature \citep[e.g., ]{Gao2004,Springel2008}.

To better understand how satellites trace the underlying dark matter, we consider all satellites that have been accreted into our host systems (including those whose stellar mass today is below $M_*=100$\msun). We find that in that case, their present-day radial distribution becomes steeper, with $\alpha_\mathrm{Ein}=0.15$, closer to that of the underlying dark matter (light green curve on left panel of Fig.~\ref{fig:radial_dist}). This result can be understood in terms of stripping: objects in tightly bound orbits suffer intense tidal disruption that bring their mass below our ``detection" threshold of $M_*=100$\msun. Such satellites are still tracked by our method and, when considered, they are faithful tracers of the underlying dark matter distribution, consistent with \cite{Green2021}. However, when observational constraints as a minimum stellar mass is included, it results on a missing subpopulation of objects that have been more stripped, which happens mostly in the inner regions of groups and clusters, leading to the biased radial distribution. 

We explore this stellar mass dependency further on the right panel of Fig.~\ref{fig:radial_dist}. Satellites are divided in 3 mass bins: massive ($M_*>10^9$\msun, orange), classic dwarf satellites ($M_*=[10^5 \rm - 10^9]$\msun, blue) and ultra-faint dwarfs ($M_*=[10^2 \rm - 10^5]$\msun, magenta).  The radial distribution of ultra-faint satellites is normalized to match the dark matter profile at the virial radius, while the distributions for the other two stellar mass bins are scaled relative to the ultra-faints based on the total number of satellites in each mass range. The error bars represent Poisson uncertainties. 

We observe that the satellites act as increasingly biased tracers of the dark matter with an increase in their stellar mass. We find $\alpha_\mathrm{Ein}=0.26, 0.49$ and $0.46$ for ultra-faint, classic dwarf and massive satellites respectively, compared to $\alpha \sim 0.17$ describing the underlying dark matter in the host. This is likely the effect of tidal stripping moving objects to the next fainter bin preferentially in the inner regions, resulting on increasingly cored radial satellite distributions when a minimum $M_*$ is imposed. We highlight the contributions from Type-1 and -2 satellites using dashed and dotted thin colored lines, respectively. Note that the relevance of the modeling of Type-2 objects increases sharply with decreasing stellar mass (and below $M_* \sim 10^9$\msun) and in particular in the inner regions. The region where Type-2 satellites dominate the population increases from never (for the brightest bin) to $r/r_{200}\sim 0.15$ for classical dwarfs to $r/r_{200}\sim 0.4$ in the case of ultra-faints. 

\section{Conclusions}
\label{sec:conclusions}

We extend the predictions of cosmological hydrodynamical simulations by explicitly modeling the tidal evolution of unresolved galaxies using analytical tidal evolution. In particular, we use the analytical and empirical results presented in \citetalias{Errani2021, Errani2022} to augment the power of the cosmological hydrodynamical simulation TNG50 into the ultra-faint galaxies regime.

The model takes advantage of the results taken directly from the hydrodynamical simulation for galaxies considered well resolved ($M_* > 5 \times 10^{6}$\msun\; for mass convergence and $M_* > 10^{8}$\msun\; for size convergence) but tracks the infall, orbit and subsequent evolution of unresolved subhalos, which are later populated with a stellar mass and stellar size modelled after extrapolations of observational abundance-matching and mass-size results. The tidal evolution after infall into the clusters is later modeled analytically for these objects. This allows us to make predictions for the stellar mass and radial distribution of satellite galaxies  down to $M_* = 100$\msun\; in host galaxy clusters with $M_{200} \sim 10^{14}$\msun, extending the range where the simulation provides reliable results by $\sim 5$ dex in stellar mass. Our method allows us to alleviate numerical issues related to resolution and artificial disruption of satellites, highlighting an exciting prospect for the discovery of ultra-faint dwarfs in nearby clusters like Fornax and Virgo. The main results of our work can be summarized as follows.  

\begin{enumerate}

\item Assuming a nearly exponential light profile for each dwarf satellite, tidal stripping evolves galaxies mostly along the stellar mass - size relation after infall into high-density environments. Tidal evolution affects more rapidly the stellar content of the galaxies than their size. We therefore predict $M_*$ - size relations for satellite dwarfs that are close in median and scatter to those observed for dwarfs in the Local Group. 

\item Depending on uncertainties on the specific abundance matching relation that ultra-faint galaxies follow, we expect $2,000$-$7,000$ satellites with $M_* \geq 100$~\msun\; within the virial radius of such hosts. Moreover, subtleties in the shape of the satellite stellar mass function below $M_* \sim 10^6$~\msun, may help shed light on the mapping between stellar mass and halo mass at infall, providing an avenue to inform abundance matching relations themselves. 
 
\item Completeness calculations indicate that sensitivities $\Sigma > 28$~mag/arcsec$^2$, comparable to upcoming data from Vera Rubin and the Roman Observatory, will be able to map satellite mass functions complete at 90\% level down to $M_*>10^5$\msun\; and 50\% level for $M_* \sim 1000$~\msun, unlocking the study of a few hundred to thousands ultra-faint dwarfs per system in richer environments than those in the Local Group. 

\item While we find that the remnants of all satellites that ever infall into a cluster follow the underlying density profile of the dark matter, imposing any luminosity or stellar mass cut may result on the prediction of more cored radial distribution by removing the most tidally affected objects that typically populate the inner few kiloparsec of each host system. For satellites with $M_*>10^6$\msun, the radial distribution resembles an Einasto profile with $\alpha_\mathrm{Ein}=0.5$, larger (and therefore shallower) than that of the dark matter. For the ultra-faint population, the inclusion of objects as faint as $M_* \sim 100$\msun\; allows to retain some more of the tidally affected objects, steepening the radial distribution to $\alpha_\mathrm{Ein} \sim 0.26$, closer to the $\alpha_\mathrm{Ein} \sim 0.2$ expected for the dark matter halo in the host. We therefore conclude that the distribution of satellite galaxies can be used as a powerful test of $\Lambda$CDM, in particular, as we reach fainter and fainter populations. 
\end{enumerate}

While we demonstrated the use of this approach to make predictions for massive galaxy clusters, this can easily be extended to the less massive systems such as groups or even individual galaxies such as Cen-A, the Milky way or M31-mass systems \citep[see ][for a similar approach to model MW-like satellites]{SantosSantos2025}. Our results in the Virgo-like cluster regime highlight the potential for group and cluster environments to inform cosmology and galaxy formation models once the numerical limitations of cosmological simulations are carefully addressed.

\section*{Acknowledgements}
PS, LVS and JAB are grateful for partial financial support from NSF-CAREER-1945310, NSF-AST-2107993 and NSF-AST-2408339 grants. LVS acknowledges the hospitality of the Max-Planck Institute for Astrophysics, where some of this work was completed. This research was supported in part by grant NSF PHY-2309135 to the Kavli Institute for Theoretical Physics (KITP). Computations were performed using the computer clusters and data storage resources of the HPCC, which were funded by grants from NSF (MRI-2215705, MRI-1429826) and NIH (1S10OD016290-01A1). We digitised the plots of \cite{Errani2022} using Automeris tool \cite{Rohatgi2022}. We thank Andrey Kravtsov and Simon White for beneficial discussions that helped improve the paper. 

\section*{Data Availability}

This paper is based on halo catalogs and merger trees from the Illustris-TNG Project (Nelson et al. 2019a,b). These data is publicly available at https://www.tng-project.org/. Catalog of galaxies resulting from our hybrid model will be shared upon request to the corresponding author if no further conflict exists with ongoing projects.



\bibliographystyle{mnras}
\bibliography{refs} 




\appendix

\section{Testing the model}
\label{sec:validation}

A validation of our model can be made using a subset of sufficiently resolved satellites for which we have predictions from the hydrodynamical simulation at $z=0$ which can be directly compared to those we would predict using our tidal evolution model. We therefore pick a subset of subhalos tracked by Subfind until $z = 0$ from FoF-0 which have a maximum stellar mass in a range $10^{8 - 9.5}$\msun. These subhalos are resolved with $\sim 1000 \rm - 50,000$ stellar particles and $\sim 10^6$ dark matter particles, providing sufficient resolution to track their evolution. They are also dark matter dominated objects, with typical contribution of the stars to the total mass $\sim 15\%$ (measured within twice the half-mass radius of the stars, $2r_{h,*}$). We take the input properties for our subset at infall time, as described in Section~\ref{ssec:input_extraction}, and we evolve the subhalos until $z = 0$ using our analytical model. 

\begin{figure*}
    \centering
    \includegraphics[width=1\linewidth]{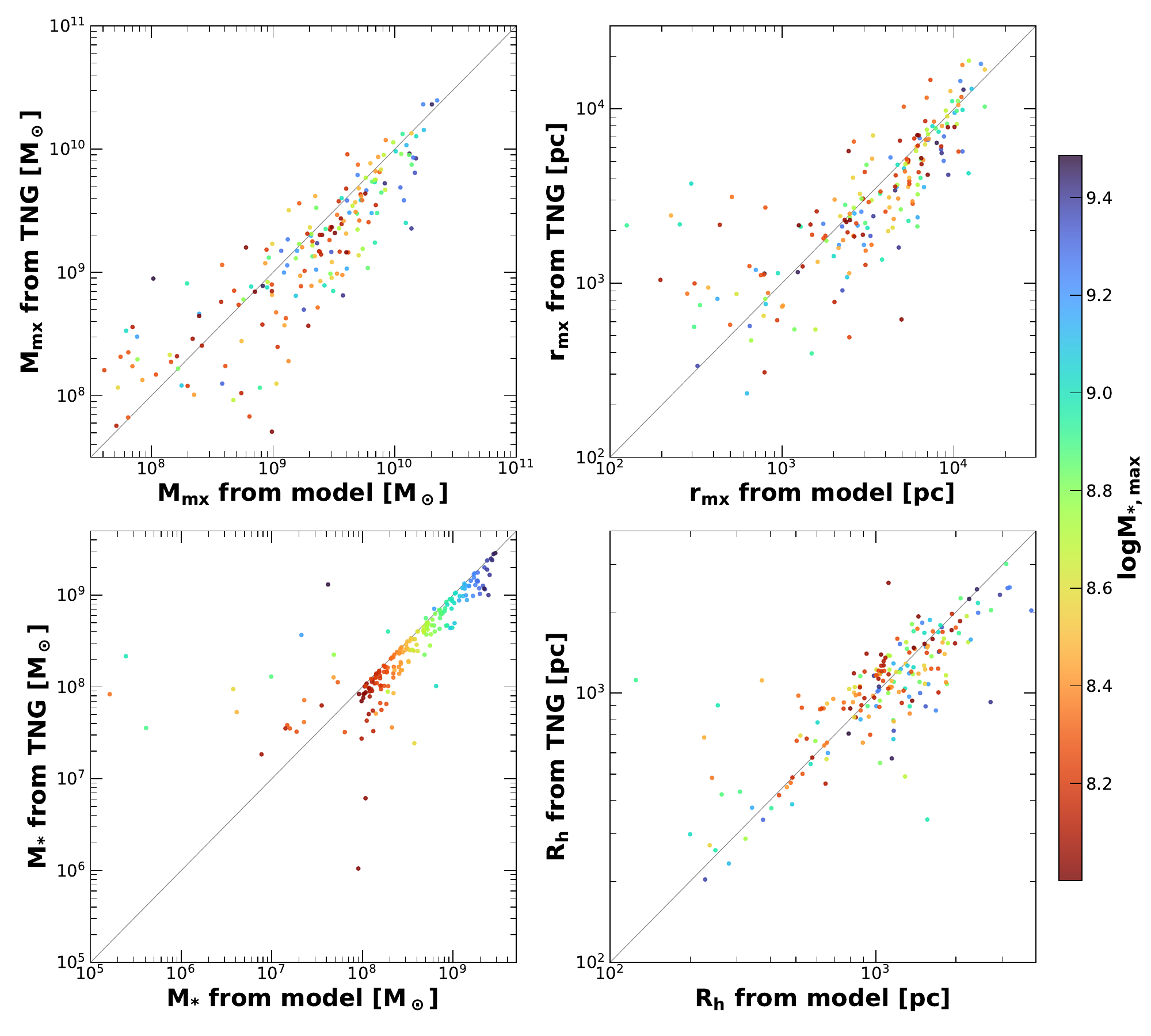}
    \caption{A validation of the model by comparing properties of subhalos at $z = 0$ between TNG50 simulation and the model for Type-1 subhalos with a maximum stellar mass post-infall $M_\mathrm{*, max}$ between $10^{8-9.5}$\msun. The points are colored by their $M_{*,\rm max}$. In each of the panels, gray solid line represents 1:1 line which would be the case when simulation and model results match exactly. With some scatter, the predictions from the model and the simulation are in good mutual agreement, suggesting that the model is capable of following the evolution of subhalos reliably within the scatter.}
    \label{fig:comparison}
\end{figure*}

We show the comparison between our model and the predictions of the simulation for this subset of objects in Fig.~\ref{fig:comparison}. The top row illustrates the dark matter mass and size expected for the satellites at $z=0$, here parametrized by the radius of maximum circular velocity, $r_{\rm mx}$ (top right), along with the mass enclosed within it, or $M_{\rm mx}$ (top left). Similarly, the bottom row compares the predictions for the present-day stellar mass, $M_*$ and size (half light radius, $R_h$) in the left and right panels, respectively. Individual satellites are shown with symbols and color-coded according to their maximum stellar mass. 

In general, there is a good correlation between values taken directly from the simulation (vertical axes) with those predicted by the model (horizontal axes), albeit with some appreciable scatter. Dashed black lines show a 1:1 relation, which roughly describes the average behavior of the points in all panels. The lack of correlation between deviations from the 1:1 relation with $M_{*,\rm max}$ is also a sign that no systematic trend with resolution is further identified here. The tight relation between stellar mass in model and simulation signals negligible stripping of stars in majority of these Type-1 cases analyzed here. This fact has been previously reported in studies such as \cite{Santos-Santos2022}. Discrepancies in estimated stellar mass by the model deviates from that of the simulation in the low mass end in the extremely stripped cases which are the outliers with larger scatter in this plot.  

Several factors can contribute to explain the scatter seen between simulation and model quantities. For example, all satellites are assumed to follow an exponential 3D energy distribution for their stellar component, which does not necessarily apply to each individual satellite in the simulation (although it does agree with the {\it average} profile for the sample. Second, the model assumes no changes in structure after infall, while the simulation may continue forming stars for 1 or 2 Gyr after infall \citep[see e.g., ][]{RodriguezGomez2015}. While we account for this by using $M_{*, \rm max}$ instead of the instantaneous mass at infall, fluctuations in the structure of the stellar component are less easy to account for. By default, we consider the size of the stellar component at the time of $M_{*, \rm max}$ and consider their evolution by stripping only afterwards. 

Overall, given the possible sources of uncertainties and scatter, the level of agreement between the model and the simulations for the {\it resolved} population is reassuring and provides an estimate of the level of variation that is to be expected in posterior predictions of our model.  

\bsp	
\label{lastpage}
\end{document}